\documentclass[]{article}

\usepackage{blindtext} 

\usepackage[sc]{mathpazo} 
\usepackage[T1]{fontenc} 
\usepackage{microtype} 
\usepackage{graphicx}
\usepackage[colorlinks=true,linkcolor=blue,citecolor=blue]{hyperref}

\usepackage[english]{babel} 

\usepackage[hmarginratio=1:1,top=32mm,columnsep=20pt,margin=0.75in]{geometry} 
\usepackage[hang, small,labelfont=bf,up,textfont=it,up]{caption} 
\usepackage{booktabs} 

\usepackage{lettrine} 

\usepackage{enumitem} 
\setlist[itemize]{noitemsep} 

\usepackage{abstract} 

\usepackage{titlesec} 
\titleformat{\section}[block]{\large\scshape\centering}{\thesection.}{1em}{} 
\titleformat{\subsection}[block]{\large}{\thesubsection.}{1em}{} 

\usepackage{fancyhdr} 
\usepackage{titling} 

\usepackage{hyperref} 
\usepackage[numbers]{natbib}
\usepackage{bm}
\usepackage{amsmath,amssymb}
\usepackage{xcolor}
\usepackage{xspace} 
\usepackage{multirow}

\title{\bfseries Personalized and uncertainty-aware coronary hemodynamics simulations: From Bayesian estimation to improved multi-fidelity uncertainty quantification \vspace{-1em}} 

\author{
Karthik Menon \textsuperscript{1, 2}, Andrea Zanoni \textsuperscript{1, 2}, Owais Khan\textsuperscript{3}, Gianluca Geraci \textsuperscript{4},\\ Koen Nieman \textsuperscript{5, 6}, Daniele E. Schiavazzi \textsuperscript{7}, Alison L. Marsden \textsuperscript{1, 2, 8} \vspace{1ex} \\
\textit{\normalsize \textsuperscript{1} Department of Pediatrics (Cardiology), Stanford School of Medicine, Stanford, CA, USA} \\
\textit{\normalsize \textsuperscript{2} Institute for Computational and Mathematical Engineering, Stanford University, Stanford, CA, USA} \\
\textit{\normalsize \textsuperscript{3} Department of Electrical, Computer, and Biomedical Engineering, Toronto Metropolitan University, Toronto, ON, Canada} \\
\textit{\normalsize \textsuperscript{4} Center for Computing Research, Sandia National Laboratories, Albuquerque, NM, USA} \\
\textit{\normalsize \textsuperscript{5} Division of Cardiovascular Medicine, Stanford School of Medicine, Stanford, CA, USA} \\
\textit{\normalsize \textsuperscript{6} Department of Radiology, Stanford School of Medicine, Stanford, CA, USA} \\
\textit{\normalsize \textsuperscript{7} Department of Applied and Computational Mathematics and Statistics, University of Notre Dame,
Notre Dame, IN, USA} \\
\textit{\normalsize \textsuperscript{8} Department of Bioengineering, Stanford University, Stanford, CA, USA}
\vspace{1em}\\
\normalsize Email for correspondence: karthikmenon@stanford.edu
\vspace{-1em}
}

\date{} 

\renewenvironment{abstract}
 {\quotation\small\noindent\rule{\linewidth}{.5pt}\par\smallskip
  {\centering\bfseries\abstractname\par}\medskip}
 {\par\noindent\rule{\linewidth}{.5pt}\endquotation}

\renewcommand{\maketitlehookd}{
\renewcommand{\abstractname}{\vspace{-1em}} 
\begin{abstract}
\normalsize
\noindent 
\noindent \textbf{Background:} Non-invasive simulations of coronary hemodynamics have improved clinical risk stratification and treatment outcomes for coronary artery disease, compared to relying on anatomical imaging alone. However, simulations typically use empirical approaches to distribute total coronary flow amongst the arteries in the coronary tree, which ignores patient variability, the presence of disease, and other clinical factors. Further, uncertainty in the clinical data often remains unaccounted for in the modeling pipeline.\\[1ex]
\noindent \textbf{Objective:} We present an end-to-end uncertainty-aware pipeline to (1) personalize coronary flow simulations by incorporating vessel-specific coronary flows as well as cardiac function; and (2) predict clinical and biomechanical quantities of interest with improved precision, while accounting for uncertainty in the clinical data.\\[1ex]
\noindent \textbf{Methods:} We assimilate patient-specific measurements of myocardial blood flow from clinical CT myocardial perfusion imaging to estimate branch-specific coronary artery flows. Simulated noise in the clinical data is used to estimate the joint posterior distributions of the model parameters using adaptive Markov Chain Monte Carlo sampling. Additionally, the posterior predictive distribution for the relevant quantities of interest is determined using a new approach combining multi-fidelity Monte Carlo estimation with non-linear, data-driven dimensionality reduction. This leads to improved correlations between high- and low-fidelity model outputs. \\[1ex]
\noindent \textbf{Results:} Our framework accurately recapitulates clinically measured cardiac function as well as branch-specific coronary flows under measurement noise uncertainty. We observe substantial reductions in confidence intervals for estimated quantities of interest compared to single-fidelity Monte Carlo estimation and state-of-the-art multi-fidelity Monte Carlo methods. This holds especially true for quantities of interest that showed limited correlation between the low- and high-fidelity model predictions. In addition, the proposed multi-fidelity Monte Carlo estimators are significantly cheaper to compute than traditional estimators, under a specified confidence level or variance.\\[1ex] 
\noindent \textbf{Conclusions:} The proposed pipeline for personalized and uncertainty-aware predictions of coronary hemodynamics is based on routine clinical measurements and recently developed techniques for CT myocardial perfusion imaging. The proposed pipeline offers significant improvements in precision and reduction in computational cost.
\end{abstract}
}

\newcommand\numberthis{\addtocounter{equation}{1}\tag{\theequation}}
\newcommand{\mpi}{MPI\textsubscript{CT}\xspace}
\newcommand{\params}{\boldsymbol{\theta}\xspace}
\newcommand{\paramshf}{\boldsymbol{\theta}^{3D}\xspace}
\newcommand{\paramslf}{\boldsymbol{\theta}^{0D}\xspace}
\newcommand{\latenthf}{\boldsymbol{z}^{3D}\xspace}
\newcommand{\latentlf}{\boldsymbol{z}^{0D}\xspace}
\newcommand{\disthf}{\pi^{3D}}
\newcommand{\distlf}{\pi^{0D}}
\newcommand{\dimlf}{{d^{0D}}\xspace}
\newcommand{\dimhf}{{d^{3D}}\xspace}
\newcommand{\dimlatenthf}{{d_r^{3D}}}
\newcommand{\dimlatentlf}{{d_r^{0D}}}
\newcommand{\Nhf}{{N_{3D}}\xspace}
\newcommand{\Nlf}{{N_{0D}}\xspace}
\newcommand{\Qhf}{Q^{3D}\xspace}
\newcommand{\Qlf}{Q^{0D}\xspace}
\newcommand{\Qmc}{\widehat{Q}^{MC}}
\newcommand{\Qmfmc}{\widehat{Q}^{MFMC}}
\newcommand{\Qmfmcae}{\widehat{Q}^{MFMC,AE}}
\newcommand{\var}{\mathbb{V}\text{ar}}
\newcommand{\cov}{\mathbb{C}\text{ov}}
\newcommand{\encoderhf}{\mathcal{E}^{3D}}
\newcommand{\decoderhf}{\mathcal{D}^{3D}}
\newcommand{\encoderlf}{\mathcal{E}^{0D}}
\newcommand{\decoderlf}{\mathcal{D}^{0D}}
\newcommand{\normflowhf}{\mathcal{T}^{3D}}
\newcommand{\normflowlf}{\mathcal{T}^{0D}}

\begin{document}

\maketitle

\section{Introduction}
\label{sec:intro}

Patient-specific computational modeling of coronary blood flow is an emerging technique for non-invasive diagnosis and treatment planning for coronary artery disease (CAD)~\citep{Menon_Hu_Marsden_2024, TAYLOR2023116414}. The use of patient-specific computational fluid dynamics (CFD) simulations to guide treatment planning for CAD has resulted in improved diagnostic accuracy and fewer unnecessary invasive procedures compared to clinical decisions guided by anatomical imaging alone~\citep{Koo2011DiagnosisNoni,Min2012,Karady2020}. Computational modeling has also been used to correlate disease severity and progression with biomechanical stimuli which are often inaccessible from imaging alone~\citep{Morbiducci2007HelicalStudy,Ramachandra2015ComputationalPressure,Ramachandra2016Patient-SpecificGrafts,Ramachandra2017GradualRemodelling,Khan2020LowGrafting,Ramachandra2022BiodegradableModel,Candreva2022RiskAngiography,Menon2023}. 

A crucial element in using computational models to guide clinical decisions is their personalization based on available patient-specific clinical data. However, there remain two major challenges in this context. First, current models are personalized primarily in terms of the anatomy, i.e., simulations use anatomical models constructed from patient-specific clinical imaging data. While additional personalization has been explored, generally in the form of tuning boundary conditions to match gross hemodynamic measurements of total flow and/or blood pressure, this does not identify personalized flow splits to each vessel in the coronary tree. Instead, flow splits are generally prescribed using empirical Murray's law-based relationships between vessel diameters and the associated flow~\citep{Murray1926,Zhou1999}. This does not account for vascular disease and variability amongst individuals. This deficiency was highlighted in previous work by Menon \textit{et al.}~\cite{Menon2024} and Xue~\textit{et al.}~\citep{Xue2023PersonalizedReserve}. These studies achieved personalized coronary flow distributions by leveraging dynamic stress CT myocardial perfusion imaging (\mpi), a non-invasive imaging technique to quantify the distribution of blood flow in the myocardium~\citep{Nieman2020DynamicImaging}. Specifically, Menon~\textit{et al.} showed significant differences in coronary flows, and both studies showed differences in \emph{fractional flow reserve} (FFR) \citep{Pijls1996} -- invasive measurements of the latter being the gold-standard for the clinical assessment of CAD -- between models based on vessel-specific personalization versus the standard diameter-based distribution of flow amongst coronary arteries. This emphasizes the importance of more fine-grained personalization to accelerate the reliable and accurate adoption of hemodynamic models in the clinic.  

Second, while the assimilation of clinical data into personalized computational models is promising, it introduces another challenge. Namely, models that are personalized based on clinical measurements usually do not account for the uncertainty in this clinical data. The assimilation of noisy data into computational models naturally induces uncertainty in the model parameters and output quantities of interest (QoIs)~\citep{NINOS2021106021}. Previous work to address this challenge has largely focused either on the estimation of probabilistic model parameters based on noisy clinical data, or on the propagation to probabilistic hemodynamic predictions, assuming known parameter distributions. The former has been performed by combining model reduction with Bayesian estimation through adaptive Markov chain Monte Carlo (MCMC)~\citep{Tran2017,SALVADOR2023107402,richter2024bayesianwindkesselcalibrationusing}. The primary hurdle in the latter has been the prohibitive computational cost of the large number of expensive model evaluations required to estimate uncertain QoIs with reasonable precision. This has previously been tackled with polynomial chaos-based propagation~\citep{Eck2016, Boccadifuoco2018ImpactAneurysms, Tran2019UncertaintyGrafts}, where the degree of success strongly depends on the problem's dimensionality and the smoothness of the underlying stochastic response. More recently, multi-level and multi-fidelity approaches for uncertainty propagation~\citep{Peherstorfer2016OptimalEstimation} have shown encouraging variance reduction in predicted QoIs for high-dimensional problems. These methods combine outputs from a few expensive high-fidelity simulations with a large number of cheap low-fidelity simulations, and the variance reduction directly depends on the correlation between low- and high-fidelity model outputs~\citep{Fleeter2020MultilevelHemodynamics,Seo2019Multi-fidelityUncertainty}. However, depending on assumptions relating to modeling the physics of the problem in the low- and high-fidelity contexts, possibly dissimilar parametrization between these models, etc., this correlation may be limited. Moreover, only a few studies have developed end-to-end, uncertainty-aware pipelines spanning clinical data to model predictions (see, e.g., a study in congenital heart disease \citep{Schiavazzi2016}), particularly for CAD~\citep{Seo2019Multi-fidelityUncertainty}.

This work combines novel clinical imaging and probabilistic modeling to address two challenges: (1) the estimation of patient-specific and \emph{vessel-specific} coronary flows based on clinical \mpi, under uncertainty; and (2) computation of posterior predictive clinical and biomechanical indicators using a new multi-fidelity approach for uncertainty propagation that leverages non-linear dimensionality reduction to better correlate models and, in turn, increase estimator confidence. One significant contribution of this work is the development of an end-to-end pipeline for personalized quantification of coronary hemodynamics based on both routine and CT perfusion-based clinical measurements that is able to account for cardiac function as well as vessel-specific coronary flow. A second key contribution is the demonstration of multi-fidelity uncertainty propagation combined with non-linear dimensionality reduction to significantly improve prediction confidence for coronary hemodynamics simulations. The method presented here significantly extends previous efforts relying on linear dimension reduction strategies for multi-fidelity sampling~\citep{Geraci2018,Geraci2018b,Fleeter2020MultilevelHemodynamics,ZengGeraci2023} by generalizing the recent work of Zanoni \textit {et al.}~\citep{ZANONI2024117119,SciTech,ArxivNeurAM}, and investigates its application to personalized coronary hemodynamics. This framework is especially useful for applications where well-correlated low-fidelity models are unavailable, and consequently, the current state-of-the-art in multi-fidelity uncertainty quantification does not yield significant reductions in variance or computational cost. We will show that this is true for various important clinical and biomechanical hemodynamic metrics, and that the proposed framework leads to significantly lower computational cost in uncertainty-aware model predictions. 

\section{Methods}
\label{sec:methods}

\begin{figure}
\centering
\includegraphics[scale=1.0]{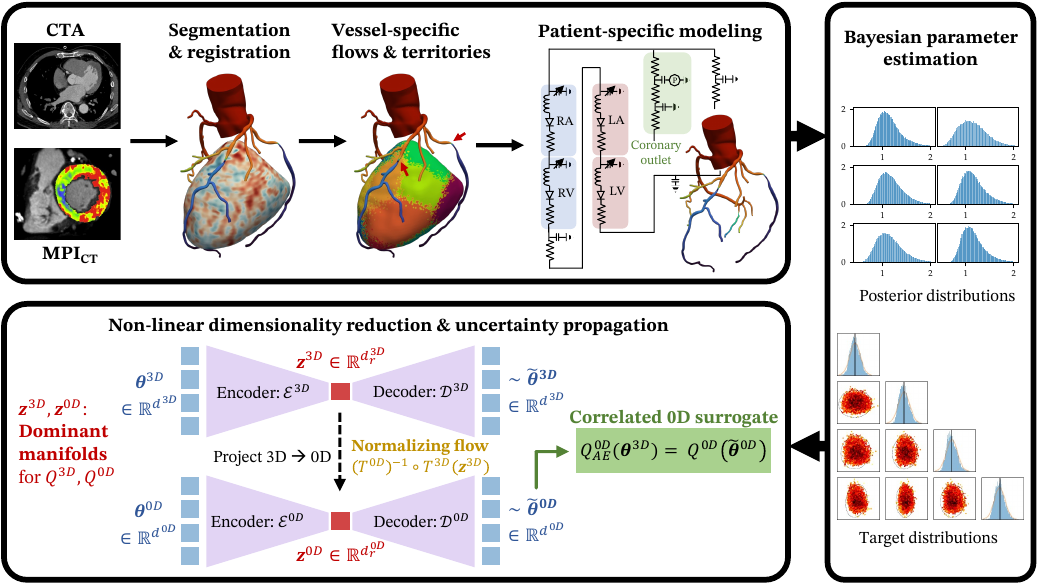}
\caption{An overview of the pipeline developed in this work, from clinical image analysis, to Bayesian parameter estimation, and, finally, the computation of posterior predictive quantities of interest.}
\label{fig:workflow}
\end{figure}

In this section, we describe the framework developed in this work. We discuss the clinical data employed, the computational modeling techniques for coronary flows, the estimation of personalized model parameters, and finally uncertainty propagation for clinical and biomechanical QoIs. 
An overview of the entire pipeline is provided in Figure~\ref{fig:workflow}.

\subsection{Clinical imaging and patient data}
\label{sec:clinical_data}

We constructed a patient-specific three-dimensional coronary artery anatomical model from coronary computed tomography angiography (CCTA) for a patient with coronary artery disease who underwent CCTA and CT myocardial perfusion imaging (\mpi) at Stanford University School of Medicine, Stanford CA, USA. The imaging was part of an ongoing study (NCT03894423) that was approved by the Institutional Review Board at the Stanford University School of Medicine. Written informed consent was received prior to patient participation. Dynamic stress CT myocardial perfusion imaging (\mpi) and CCTA were performed on a third-generation dual-source CT scanner (SOMATOM Force, Siemens Healthineers) following a standard protocol \citep{Mahnken2010QuantitativeExperience,Nieman2020DynamicImaging} with images acquired at systole and mid-diastole, respectively. The CCTA and \mpi images were co-registered using an affine registration method in the open-source \textit{3D Slicer} \textit{(www.slicer.org)} software. Details on the imaging and registration protocol are available in Menon \textit{et al.}~\cite{Menon2024} and Khan \textit{et al.}~\citep{KhanQuantification2024}. We then used the open-source \textit{SimVascular} software package to segment and reconstruct a three-dimensional patient-specific model of the coronary artery anatomy from the CCTA imaging. The left ventricle (LV) myocardium was segmented from the \mpi images using thresholding in \textit{ParaView} \textit{(www.paraview.org)}. This produced co-registered three-dimensional models of the coronary tree and LV myocardium (top panel of Figure~\ref{fig:workflow}). 

We then used the coronary tree anatomical model, segmented LV volume and measured myocardial blood flow (MBF) from \mpi to determine the flow corresponding to each coronary artery branch. Note that not all the coronary arteries perfuse the LV (where MBF data was collected). Therefore, we only computed branch-specific flows for those branches perfusing the LV at this stage. The patient was right-dominant and the arteries perfusing the LV were determined in consultation with an expert clinician. We first performed a Voronoi tessellation on the LV volume, where each point in the LV was assigned to its closest coronary artery outlet. This resulted in the LV being divided into non-overlapping sub-volumes corresponding to each coronary artery branch, which physiologically represent the LV perfusion volume of each coronary artery (see top panel of Figure~\ref{fig:workflow}). This method is in line with several prior studies which have based LV perfusion territory selections on the unweighted or weighted distance of each point on the LV to either the closest coronary outlet or the coronary centerline \citep{DiGregorio2021AMicrovasculature,Papamanolis2021}. We showed in previous work that our personalization procedure is robust to the choice of method to compute branch-specific LV perfusion territories~\citep{Menon2024}. Following this, the volumetric flow-rate associated with each coronary artery branch was determined by integrating the MBF within the corresponding sub-volume of the myocardium.

To account for coronary flow uncertainty from the imaging and analysis described above, we introduced two sources of noise into the procedure. One was in the computation of the LV perfusion territories corresponding to each coronary artery, where we introduced noise into the distance between each point on the LV and the coronary artery outlets. This simulated anatomical uncertainty in the locations of the arteries and the shape of the LV myocardium stemming from the segmentation and registration procedures outlined above. We specified Gaussian noise with zero mean and a standard deviation of 10\% of the baseline (deterministic) distance. The second source of noise relates to the MBF data, and accounts for uncertainty in the underlying image acquisition and analysis that is used to compute MBF from dynamic contrast imaging of the LV \cite{Mahnken2010QuantitativeExperience}. This was specified as Gaussian noise with zero mean and a standard deviation of 20\% of the baseline (clinical) MBF. We note that these noise levels were chosen based on prior experience working with this data, due to the absence of repeated clinical measurements of \mpi and its variability. However, clinically-informed noise distributions could be easily incorporated into the current framework and would provide more realism. We computed statistics by simulating 2500 realizations of the branch-specific MBF and LV perfusion volume tessellation. This yielded uncertainty-aware distributions for the measured flow corresponding to each coronary branch.

\subsection{Computational modeling}
\label{sec:modeling}

The multi-fidelity framework developed in this work relied on a combination of high-fidelity (HF) three-dimensional (3D) simulations of coronary artery flow combined with low-fidelity (LF) zero-dimensional (0D) flow models. For the three-dimensional patient-specific simulations, we used the \textit{svSolver} flow solver, which is a part of the \textit{SimVascular} software suite. \textit{svSolver} combines a stabilized finite element spatial discretization using linear tetrahedral elements with a second-order accurate generalized-$\alpha$ time stepping~\citep{Jansen1999,Whiting2001ABasis} scheme. The governing equations are the three-dimensional incompressible Navier-Stokes equations, 
\begin{equation}
    \nabla \cdot \boldsymbol{u} = 0 \;\; ; \;\; \rho \frac{\partial \boldsymbol{u}}{\partial t} + \rho \boldsymbol{u} \cdot \nabla \boldsymbol{u} = -\nabla p + \mu \nabla^2 \boldsymbol{u},
    \label{eq:navier-stokes}
\end{equation}
where $\boldsymbol{u}$ and $p$ are the blood velocity and pressure, respectively. Blood was assumed to be Newtonian with density $\rho = 1.06$ g/cm$^3$ and viscosity of $\mu = 0.04 $ dynes/cm$^2$. Note that non-Newtonian effects begin to appear for blood vessels with diameters below 300 $\mu$m~\citep{Secomb2017BloodMicrocirculation}, which was well below the size of the vessels modeled in this study. For simplicity, we treated coronary artery walls as rigid, and mesh convergence was assessed using a similar coronary flow modeling setup in prior work~\citep{Menon2023}. Each 3D simulation consisted of 5 cardiac cycles, with 1000 time-steps for each cycle, requiring approximately 7 hours on 96 processors of a high performance computing cluster and approximately 1 hour for postprocessing on a single processor. 

Zero-dimensional flow simulations were performed using \textit{svZeroDSolver}, an open-source solver for lumped parameter (0D) hemodynamic simulation, provided with the \textit{SimVascular} software suite. Zero-dimensional models approximate vascular hemodynamics using an electric circuit analogy, specifically, using resistors to model major viscous losses, capacitance to model vessel compliance, and inductance to represent blood flow inertia. 0D model parameters were estimated based on the 3D anatomy (i.e., coronary branch diameters and lengths) using an automated procedure described in Pfaller \textit{et al.}~\citep{Pfaller2022AutomatedFlow}. Each zero-dimensional simulation was run for 5 cardiac cycles, with 1000 time-step per cycle, and took approximately 10 seconds, including post-processing, on a single processor.  

Boundary conditions at the inlet and outlet of the aorta and coronary outlets in both 3D and 0D simulations were prescribed through a closed-loop 0D representation of the systemic circulation and distal vascular resistance. In particular, we used Windkessel boundary conditions at the aortic outlet~\citep{Vignon-Clementel2006} and modeled the four heart chambers in the systemic circulation using time-varying elastance. Coronary outflow boundary conditions included the effect of distal resistance as well as the intra-myocardial pressure, which is essential to capture the diastolic nature of coronary flow~\citep{Kim2010}. Flow simulations were performed under hyperemic conditions by scaling the resistance distal to coronary arteries by 0.24 compared to their baseline at rest~\citep{Taylor2013}. This was done to match the acquisition of \mpi under hyperemia. The estimation of the boundary condition parameters, based on matching clinical measurements, is described in Section~\ref{sec:estimation}. The top panel of Figure~\ref{fig:workflow} shows a schematic of the computational model.

\subsection{Parameter estimation using Markov Chain Monte Carlo}
\label{sec:estimation}

The parameters of the closed-loop 0D boundary conditions were estimated so that the simulations recapitulated clinically measured patient-specific data. This was performed in two stages. In the first stage, we estimated the total systemic and coronary resistance and capacitance, the elastance and volumes of the heart chambers, cardiac activation times and intramyocardial pressure so that the simulations produced satisfactory agreement with the clinically measured systolic and diastolic blood pressure cuff measurements, as well as LV stroke volume and ejection fraction from echocardiography. In addition, these patient-specific clinical measurements were augmented with literature-based population-averaged target data, including the waveforms for cardiac activation and coronary flow, as well as pulmonary pressure. This first stage of the estimation consisted of 36 parameters. We determined their (deterministic) point estimates using the 0D model and derivative-free Nelder-Mead optimization~\citep{Nelder1965AMinimization}. Details on the parameters, the initial guess, and optimization setup are provided in~\cite{Menon2023} and~\cite{Menon2024}.

The second stage, which was the primary focus in this work, aimed to personalize the branch-specific flow in each coronary artery based on the MBF distribution in the LV as measured by \mpi. The distribution of flow amongst the branches of the coronary artery tree is determined primarily by the distribution of outlet resistances amongst the branches. As mentioned earlier, since not all coronary arteries perfuse the LV, we focus here only on those that do. For coronary artery outlets supplying the RV or septum, the distal resistance was specified based on Murray's law with an exponent of 2.6~\citep{Murray1926,Zhou1999}. To account for the noise in the target branch-specific flows from \mpi (described in Section~\ref{sec:clinical_data}), the posterior distribution for the distal resistance at each coronary outlet perfusing the LV was determined using Bayesian estimation.

While keeping the 36 parameters optimized in the first stage fixed, we considered a set of outlet resistances $\boldsymbol{r} = \{r_1, r_2, ..., r_{N_c}\}$, where $N_c$ is the number of coronary artery outlets perfusing the LV. We generated samples from the joint probability density of $\boldsymbol{r}$, given a set of target flow rates, $\boldsymbol{f}_{CT} \in \mathbb{R}^{N_c}$ specified at each of the $N_c$ outlets. In other words, the estimation yields a sample-based characterization of the posterior distribution, $p(\boldsymbol{r} | \boldsymbol{f}_{CT})$. As in the first stage of the optimization, we used the 0D model as a surrogate to map the input parameters, $\boldsymbol{r}$, to the simulated flows at the outlets of each coronary artery, $\boldsymbol{f} \in \mathbb{R}^{N_c}$. 

According to Bayes theorem, the posterior distribution can be expressed as, 
\begin{equation}
    p(\boldsymbol{r} | \boldsymbol{f}_{CT}) = p(\boldsymbol{f}_{CT}|\boldsymbol{r}) p(\boldsymbol{r})/p(\boldsymbol{f}_{CT}).
    \label{eq:bayes}
\end{equation}
In Equation~\eqref{eq:bayes}, the likelihood $p(\boldsymbol{f}_{CT}|\boldsymbol{r})$ quantifies the ability of a given set of boundary condition parameters, $\boldsymbol{r}$, to produce model outputs matching the measured targets, $\boldsymbol{f}_{CT}$, with uncertainty induced by a known probabilistic characterization of the measurement noise. Following the discussion on how noise was added to MBF data in Section~\ref{sec:clinical_data}, each component of $\boldsymbol{f}_{CT}$ had a mean equal to the clinically measured MBF within the LV perfusion volume associated with a given coronary branch, and a standard deviation given by 20\% of the mean. Interestingly, the uncertain target coronary flows computed through the process described above were characterized by a non-diagonal covariance matrix, $\boldsymbol{\Sigma}$, which we estimated from 2500 realizations of the noisy MBF data. Note that the non-diagonal nature of the covariance comes primarily from the inter-dependence between adjacent LV perfusion territories. The likelihood is expressed as,
\begin{equation}
    p(\boldsymbol{f}_{CT}|\boldsymbol{r}) = \frac{1}{\sqrt{(2\pi)^{N_c} \det \boldsymbol{\Sigma}}} \exp{\left\{-\frac{1}{2}\left[\boldsymbol{f}_{CT} - \boldsymbol{f}(\boldsymbol{r})\right]^T \boldsymbol{\Sigma}^{-1} \left[\boldsymbol{f}_{CT} - \boldsymbol{f}(\boldsymbol{r})\right]\right\}},
    \label{eq:likelihood}
\end{equation}
where $\boldsymbol{f}(\boldsymbol{r})$ is the vector of simulated coronary outflows. Furthermore, we assumed the prior, $p(\boldsymbol{r})$ in Equation~\eqref{eq:bayes}, to be uniformly distributed with a range $[\, 0.5\, \widehat{r}, 2.0\, \widehat{r} \,]$, where $\widehat{r}$ is determined for each artery, or each component of $\boldsymbol{r}$, by simply distributing the total resistance distal to the LV-perfusing arteries (which was estimated in the first stage of parameter estimation) based on the relative target flows for each artery.

We then generated samples from the posterior distribution of personalized coronary outlet resistances, $p(\boldsymbol{r} | \boldsymbol{f}_{CT})$, using Markov Chain Monte Carlo. In particular, due to the slow convergence of the classical Metropolis-Hastings algorithm, we utilized a Differential Evolution Adaptive Metropolis (DREAM) sampler with adaptive proposal distribution~\citep{Vrugt2009AcceleratingSampling}. This method leverages self-adaptive differential evolution and runs multiple parallel Markov chains, which enhances its computational performance and convergence assessment. In this work, we ran 24 parallel chains with a total of 10,000 generations each. We used the Gelman-Rubin statistic~\citep{Gelman1992} with a threshold of 1.1 to assess convergence. We discarded 50\% of the samples as burn-in samples. Finally, at each generation we rescaled the components in $\boldsymbol{r}$ so that the total coronary resistance was preserved, consistent with the value determined at the first stage of estimation.

\subsection{Multi-fidelity uncertainty propagation}
\label{sec:propagation}

The next step of the framework was estimating the posterior predictive distribution propagating the uncertain model parameters to relevant clinical and biomechanical QoIs. 
In other words, given a vector of uncertain model parameters, $\params \in \mathbb{R}^{N_p}$, whose joint distribution was determined from the solution of the above inverse problem, we aimed to efficiently estimate the corresponding distribution for the QoI, $Q(\params): \mathbb{R}^{N_p} \rightarrow \mathbb{R}$.

In this work, $Q(\params)$ represents a relevant output from a 3D patient-specific simulation of coronary blood flow, with random inputs $\params = \{r_1, r_2, ..., r_{N_c}, s\} = \{\boldsymbol{r}, s\}$. Here, the first $N_c$ components are outlet resistances (see discussion in Section~\ref{sec:estimation}) having joint distribution equal to the posterior $p(\boldsymbol{r} | \boldsymbol{f}_{CT})$, and $s$ is a scaling factor for the total coronary resistance which accounts for the uncertainty in quantifying total coronary flow, which is a documented challenge in \mpi~\citep{Nieman2020DynamicImaging,Xue2023PersonalizedReserve}.
Assuming the coronary flow is in the range of 3\% to 5\% of the total cardiac output, we specify $s$ to be uniformly distributed with $s \sim \mathcal{U}(0.7, 1.25)$. The limits were computed to yield 3\%-5\% coronary-to-systemic flow splits and $s$ was assumed to be independent of $\boldsymbol{r}$. We denote the joint distribution of $\params = \{\boldsymbol{r}, s\}$ by $p(\params)$.

The simplest approach to quantify posterior predictive moments is through standard Monte Carlo sampling, where, for example, the mean of $Q(\params)$ can be estimated as,
\begin{equation}
    \Qmc_N = \frac{1}{N}\sum_{i=1}^{N} Q(\params_i).
    \label{eq:mc}
\end{equation}
Here, $\params_i$ is a realization of $\params \sim p(\params)$, and $N$ is the selected number of samples. This estimator is unbiased, but its variance scales as $\var[\Qmc] \sim 1/N$, leading to a computationally intractable estimation for applications, such as the current one, where each simulation is computationally expensive. This motivates our use of recently proposed multi-fidelity Monte Carlo estimators.

\subsubsection{Multi-fidelity Monte Carlo}
\label{sec:mfmc}
In this section, we briefly introduce the multi-fidelity Monte Carlo (MFMC) formulation employed in this work. Although we limit our presentation to MFMC, for simplicity of exposure, we note here that the proposed strategy is general and could be adopted with any of the existing multi-fidelity estimators recently introduced in literature; see, e.g., multi-level Monte Carlo~\citep{Giles2008,Cliffe2011,Giles2015}, multi-index Monte Carlo~\citep{Haji2016,HajiAli2016}, multi-level multi-fidelity Monte Carlo~\citep{Nobile2015,Fairbanks2017,Geraci2017}, Approximate Control Variate~\citep{Gorodetsky2020,Bomarito2022}, and multi-level best linear unbiased estimators~\citep{Schaden2020,Schaden2021,Croci2023}.

Multi-fidelity Monte Carlo estimators are designed to leverage the low computational cost of running a large number of cheap low-fidelity (LF) model evaluations to improve the variance reduction in estimating QoIs from expensive high-fidelity (HF) simulators~\citep{Ng2014,Peherstorfer2016OptimalEstimation,Peherstorfer2018}. Here, the high-fidelity and low-fidelity models are 3D and 0D coronary flow simulations, respectively, both of which are discussed in Section~\ref{sec:modeling}. We refer to QoIs computed by each of these models as $\Qhf(\params)$ and $\Qlf(\params)$, respectively. Note that in the standard multi-fidelity Monte Carlo formulation, the low- and high-fidelity models share the same parametrization, which we will discuss further in Section~\ref{sec:mfmc_ae}.

In the two-model case, i.e., where a single low-fidelity model is used as control variate, the MFMC estimator for the mean of $Q(\params)$ takes the form~\citep{Ng2014,Peherstorfer2016OptimalEstimation}
\begin{equation}
    \Qmfmc_{\Nhf,\Nlf} = \frac{1}{\Nhf} \sum_{i=1}^\Nhf \Qhf(\params_i) + \alpha \Bigg(\frac{1}{\Nlf} \sum_{i=1}^\Nlf \Qlf(\params_i) - \frac{1}{\Nhf} \sum_{i=1}^\Nhf\Qlf(\params_i) \Bigg).
    \label{eq:mfmc}
\end{equation}
Here, $\Nhf$ and $\Nlf$ are the number of 3D and 0D simulations, respectively. In particular, $\Nlf  = \Nhf + \Delta_{0D}$, where $\Nhf$ is referred to here as the pilot sample and $\Delta_{0D}$ are additional low-fidelity simulations evaluated at independently drawn samples of $\params$. It can be shown that $\Qmfmc$ is an unbiased estimator for $Q(\params)$ owing to the telescoping sum on the right hand side of Equation~\eqref{eq:mfmc}~\citep{Peherstorfer2016OptimalEstimation}.

The coefficient $\alpha$ in Equation~\eqref{eq:mfmc} is chosen to minimize the variance of the estimator and is given by~\citep{Peherstorfer2016OptimalEstimation}
\begin{equation}
    \alpha = \frac{\cov(\Qhf, \Qlf)}{\var[\Qlf]}.
\end{equation}
With this value of $\alpha$, the variance of the estimator is given by~\citep[Lemma 3.3]{Peherstorfer2016OptimalEstimation}
\begin{equation}
    \var[\Qmfmc_{\Nhf,\Nlf}] = \frac{\var[\Qhf]}{\Nhf} \Bigg[1 - \left(1 - \frac{\Nhf}{\Nlf}\right)\rho^2\Bigg] = \var[\Qmc_\Nhf] \Bigg[1 - \left(\frac{\Delta_{0D}}{\Nlf}\right)\rho^2\Bigg],
    \label{eq:mfmc_var}
\end{equation}
where $\rho$ is the Pearson's correlation coefficient between the low-fidelity and high-fidelity (0D and 3D) models. Note that this is the expression of the variance for general values of $\Nhf$ and $\Nlf$, and one can compute optimal values that minimize the variance for a given computational budget~\citep{Peherstorfer2016OptimalEstimation}. However, in this work we used a smaller number of samples to maintain a tractable computational cost. This is discussed further in Section~\ref{sec:budget}. 

Equation~\eqref{eq:mfmc_var} suggests a significant variance reduction compared to the standard Monte Carlo estimator ($\Qmc_\Nhf$) when $\Nlf \gg \Nhf$ and $\rho \approx 1$. In other words, having an inexpensive and well-correlated low-fidelity model is crucial to improve confidence in the estimates. However, this is challenging in applications where certain physical phenomena are under-resolved or even neglected by low-fidelity representations. In such a case, alternative strategies are required to construct well-correlated low-fidelity models, as discussed in the next section.

\subsubsection{Non-linear dimensionality reduction with autoencoders and normalizing flows}
\label{sec:mfmc_ae}

In this section, we describe a new approach for effectively reducing variance in multi-fidelity Monte Carlo estimates. In particular, we modify the sampling of low-fidelity models, such that their resulting outputs are well-correlated with the corresponding high-fidelity model outputs. This is accomplished by determining an optimal shared space between high- and low-fidelity representations, and performing a data-driven re-parameterization of the low-fidelity model. This method achieves higher correlation with the high-fidelity response, and is also applicable to low- and high-fidelity models having dissimilar parametrizations, generalizing classical multi-fidelity estimators, e.g.,~\cite{Geraci2018,Geraci2018b,ZengGeraci2023,ZANONI2024117119}, or surrogate-based approaches, e.g.,~\citep{Zeng2023Sci}.

We used autoencoders to identify \emph{active} low-dimensional non-linear manifolds for each QoI. Autoencoders are a data-driven approach to finding low-dimensional representations of data using neural networks. Each autoencoder consisted of a dense (fully connected) encoder $\mathcal{E}:\mathbb{R}^d \rightarrow \mathbb{R}^{d_r}$ and a dense decoder $\mathcal{D}:\mathbb{R}^{d_r} \rightarrow \mathbb{R}^d$. The encoder compresses the original $d$-dimensional data to a $d_r$-dimensional \emph{latent-space} representation, and the decoder reconstructs the $d$-dimensional data from the latent representation.

Let $\paramshf_i \sim \disthf$ and $\paramslf_i \sim \distlf$ be samples from the high-fidelity and low-fidelity random inputs, respectively, with corresponding QoIs given by $\Qhf(\paramshf): \mathbb{R}^\dimhf \rightarrow \mathbb{R}$ and $\Qlf(\paramslf): \mathbb{R}^\dimlf \rightarrow \mathbb{R}$. We trained separate autoencoders, i.e. $\encoderhf:\mathbb{R}^\dimhf \rightarrow \mathbb{R}^\dimlatenthf$, $\decoderhf:\mathbb{R}^\dimlatenthf \rightarrow \mathbb{R}^\dimhf$ for the 3D model, and $\encoderlf:\mathbb{R}^\dimlf \rightarrow \mathbb{R}^\dimlatentlf$, $\decoderlf:\mathbb{R}^\dimlatentlf \rightarrow \mathbb{R}^\dimlf$ for the 0D model, with corresponding latent variables expressed as
\begin{equation}
    \latenthf = \encoderhf(\paramshf)\,\,\text{and}\,\,\latentlf = \encoderlf(\paramslf).
    \label{eq:latentspace}
\end{equation}
Note that, unlike classical unsupervised autoencoders, which discover manifolds of reduced dimensionality directly in the input space, we instead focus on determining a latent space capturing the full variation of a given QoI for both the low- and high-fidelity model response. Therefore, we minimized a loss function of the form (shown here only for the high-fidelity model for brevity),
\begin{align*}
    \mathcal{L}^{3D}(\bm{\phi}^{3D}) &= \text{MSE}[\Qhf(\paramshf), \; \Qhf_{NN} \circ \encoderhf(\paramshf)] \\ &+ \text{MSE}[\Qhf(\paramshf), \; \Qhf_{NN} \circ \encoderhf \circ \decoderhf \circ \encoderhf(\paramshf)] \\ &+ \text{MSE}[\Qhf_{NN} \circ \encoderhf(\paramshf), \; \Qhf_{NN} \circ \encoderhf \circ \decoderhf \circ \encoderhf(\paramshf)] \\ &+ \text{MSE}[\decoderhf \circ \encoderhf(\paramshf), \; \decoderhf \circ \encoderhf \circ \decoderhf \circ \encoderhf(\paramshf)]. \numberthis
    \label{eq:loss_ae}
\end{align*}
Here, $\bm{\phi}^{3D}$ represents the trainable autoencoder parameters, and $\circ$ indicates function composition. In addition, $\Qhf_{NN}$ represents a fully-connected neural network surrogate of $\Qhf$ with inputs in the latent space, trained simultaneously with the autoencoder. We used $\Qhf_{NN}$ to circumvent the prohibitive cost of evaluating $\Qhf$ every time the loss function is computed, while also providing a supervised loss to identify the latent variables describing the function's variability. This choice is motivated by prior work Ref.~\cite{ArxivNeurAM}, though here we added the third term to reinforce the first two loss terms.

In order to map between latent spaces corresponding to the high- and low-fidelity models, we also needed to ensure that their respective latent space representations had the same probability distribution (so-called \emph{shared space}). This was achieved by estimating their probability density through normalizing flows, which determine invertible transformations, $\normflowhf$ and $\normflowlf$, from the \textit{a priori} unknown latent space probability densities to a common standard Gaussian, $\mathcal{N}(0,1)$.

We then evaluate the low-fidelity model at new sampling locations, which we expected to exhibit improved correlation with respect to $\Qhf$. To do so, every input realization for the high-fidelity model is mapped to its latent space and then the shared space. Each input is then mapped to the latent space of the low-fidelity model and finally to the input space of the low-fidelity model. Formally, this is expressed as,
\begin{equation}
    \Qlf_{AE}(\paramshf) = \Qlf \circ \decoderlf \circ (\normflowlf)^{-1} \circ \normflowhf \circ \encoderhf(\paramshf).
    \label{eq:q_lf_ae}
\end{equation}
Intuitively, we sample the low- and high-fidelity models at the same locations within the \emph{shared space}. Thus, by evaluating each model at equivalent locations within their respective influential manifolds (for a specific QoI), we expect well-correlated model outputs. Moreover, quantitative results have been shown under idealized assumptions in \cite[Proposition 4.4]{ZengGeraci2023} for the linear case, and in \cite[Theorem 3.3]{ArxivNeurAM} for the nonlinear approach presented here. Specifically, these references show that the new sampling locations for the low-fidelity model provided by the transformations in \eqref{eq:q_lf_ae} yield a correlation with respect to the high-fidelity model that is always larger than the original correlation.

In this work, we used a one-dimensional latent space, i.e., $\dimlatenthf = \dimlatentlf = 1$. For this special case, the normalizing flow can be simplified. We can convert any distribution, i.e. the arbitrarily distributed $\latenthf$ and $\latentlf$, to the standard Gaussian distribution using inverse transform sampling. Here, we used the empirical distribution function to first transform the latent space distribution to $\mathcal{U}(0,1)$ and then to $\mathcal{N}(0,1)$ by means of an invertible analytical map. 

In addition, since we aim to increase the correlation between the low- and high-fidelity models by linking them through a common shared space, we performed a second stage of training for the autoencoders prior to the final construction of $\Qlf_{AE}$. 
Here, we combined the losses from low- and high-fidelity autoencoders, and added a loss term to boost the correlation, $\rho$, between the high-fidelity model evaluated at the original samples and the \textit{resampled} low-fidelity model outputs. 
In practice, this second stage of training used a loss function of the form
\begin{equation}
    \mathcal{L}(\bm{\phi}^{3D}, \bm{\phi}^{0D}) = \mathcal{L}^{3D} + \mathcal{L}^{0D} - \Big|\rho[\Qhf, \; \Qlf_{NN} \circ (\normflowlf)^{-1} \circ \normflowhf \circ \encoderhf(\paramshf)]\Big|.
    \label{eq:loss_together}
\end{equation}
The neural network weights during this second stage were initialized with those determined from the first stage discussed above with loss function in Equation~\eqref{eq:loss_ae}.

Using the modified low-fidelity model given by Equation~\eqref{eq:q_lf_ae}, the proposed multi-fidelity estimator is expressed as
\begin{equation}
    \Qmfmcae_{\Nhf,\Nlf} = \frac{1}{\Nhf} \sum_{i=1}^\Nhf \Qhf(\paramshf_i) + \alpha \Bigg(\frac{1}{\Nlf} \sum_{i=1}^\Nlf \Qlf_{AE}(\paramshf_i) - \frac{1}{\Nhf} \sum_{i=1}^\Nhf\Qlf_{AE}(\paramshf_i) \Bigg).
    \label{eq:mfmc_ae}
\end{equation}
The estimator $\Qmfmcae$ is unbiased following the same reasoning as with the standard multi-fidelity Monte Carlo estimator in Equation~\eqref{eq:mfmc}~\citep{Peherstorfer2016OptimalEstimation}. Moreover, this was demonstrated for various analytical examples in our previous work~\citep{ZANONI2024117119}.

Note that while the framework developed here was inspired by Zanoni \textit{et al.} \citep{ZANONI2024117119,ArxivNeurAM}, we introduced several features here that significantly improved its performance. For one, while the previous work parameterized the normalizing flows using neural networks, our use of a one-dimensional latent space with the analytical transformations between probability distributions reduced the computational complexity and required dataset size. In addition, the inclusion of the correlation between low- and high-fidelity outputs in the loss, combined with the two-stage training of the autencoders, using Equations~\eqref{eq:loss_ae} and~\eqref{eq:loss_together}, significantly improved the correlation between the models as well as the reproducibility of the framework.

Details on the training of the neural networks, the hyper-parameters, and other aspects of the method are available in Appendix~\ref{app:hyperparameters}.

\section{Results}
\label{sec:results}

We now discuss the performance of the proposed data-to-prediction pipeline, from the estimation of personalized parameters to the prediction of clinical and biomechanical QoIs under clinical data uncertainty.

\subsection{Parameter estimation}
\label{sec:estimation_result}

\begin{figure}
\centering
\includegraphics[scale=1.0]{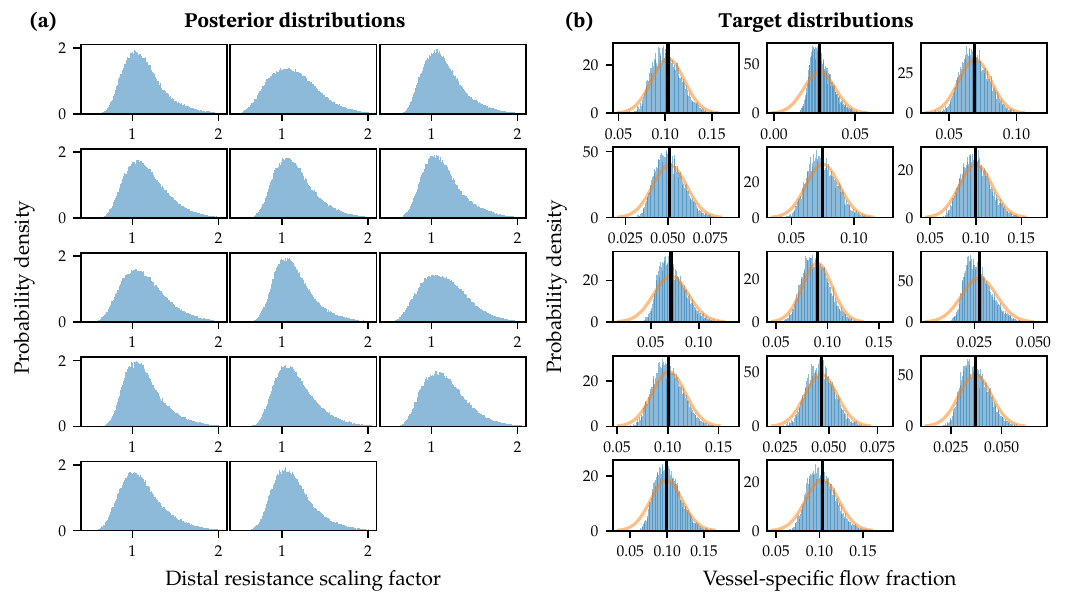}
\caption{(a) Slices of the estimated posterior distributions of the distal resistance for all 14 coronary outlets. (b) Distributions of predicted coronary flows compared with clinically measured targets (vertical line) and noise distribution for each coronary artery.}
\label{fig:dream}
\end{figure}

As described in Section~\ref{sec:estimation}, our estimation consisted of a two-step procedure -- a deterministic optimization-based step to assimilate patient-specific and literature-based measurements of cardiac function, and Bayesian estimation step to personalize vessel-specific coronary flows. Since the main focus of this work is on demonstrating the uncertainty-aware pipeline for personalized assessment of coronary hemodynamics, this section focuses on the Bayesian parameter estimation for vessel-specific coronary flows. We summarize the cardiac function estimation results in Figure~\ref{fig:cardiac_function} of Appendix~\ref{app:parameter_estimation}.

Figure~\ref{fig:dream}(a) shows the estimated marginals for the $N_c=14$-dimensional posterior distribution, $p(\boldsymbol{r})$. The unimodal character of all estimated marginals and their limited variance confirms that the estimation problem is well-posed, with identifiable resistance parameters.

In addition, Figure~\ref{fig:dream}(b) shows a comparison for the clinically-measured target coronary flow in each artery, the assumed measurement noise, and the posterior predictive marginals. Our results (in blue) closely match both the mean target flows and their distribution induced by the noise model in Equation~\eqref{eq:likelihood}. We were also able to match the covariance of the target flows, as shown in Figure~\ref{fig:results_scatter} in Appendix~\ref{app:parameter_estimation}.

\subsection{Uncertainty propagation}
\label{sec:propagation_results}

We now discuss the estimation of various clinical and biomechanical QoIs with a focus on important metrics in the diagnosis and progression of CAD. We estimated these metrics in the three primary coronary artery branches - the left anterior descending (LAD), the left circumflex (LCx), and the right coronary artery (RCA). Furthermore, these quantities of interest were computed in the vicinity of stenoses present in each of these branches (indicated by red arrows in the top panel of Figure~\ref{fig:workflow}). For each QoI, we compared the performance of the standard Monte Carlo (MC) estimator, the multi-fidelity Monte Carlo (MFMC) estimator based on the original low-fidelity model, and the proposed improved multi-fidelity Monte Carlo estimator with non-linear dimensionality reduction (MFMC-AE).

In this work, we used $\Nhf=500$ and $\Nlf=10,000$. Note that the estimated mean of each quantity of interest reported in this section was from one realization of each estimator. The confidence intervals were estimated using Chebyshev's inequality with the analytical variance, discussed in section \ref{sec:propagation}, for each estimator. Therefore, while the estimators are unbiased, the offsets seen in the mean values are expected from a single realization. In addition, for this application, we do not necessarily need a computational budget as large as our chosen $\Nhf$. We show the convergence of the method for smaller values of $\Nhf$ in Section~\ref{sec:budget}. 

\subsubsection{Lumen wall shear stress}
\label{sec:tawss_osi}

The biomechanical quantities of interest we report in this paper are the time-averaged wall shear stress (TAWSS) and oscillatory shear index (OSI). TAWSS is the time-average of the shear stress on the vessel wall at every point on the vessel wall and OSI is a metric between 0 and 0.5, which measures the degree of oscillation/direction-change in the shear-stress experienced by the vessel wall at a given point. Pathological values for both of these metrics have been correlated with the progression of vascular disease (see, e.g.,~\cite{Ku1997BloodArteries}).

From the 3D simulations, we computed the shear stress on the vessel wall as,
\begin{equation}
    \tau = \mu (\nabla\boldsymbol{u} + \nabla\boldsymbol{u}^T)\cdot\bm{n}|_{\boldsymbol{x}=\text{wall}},
\end{equation}
where $\bm{n}$ is the unit vector normal to the vessel wall at any point, and the subscript ``${\boldsymbol{x}=\text{wall}}$'' denotes evaluation at the lumen. TAWSS and OSI were then computed as,
\begin{equation}
    \text{TAWSS} = \int_0^T \tau \; dt,
\end{equation}
\begin{equation}
    \text{OSI} = \frac{1}{2}\Bigg(1-\frac{|\int_0^T \tau dt|}{\int_0^T |\tau| dt}\Bigg).
\end{equation}
Here, $T$ denotes the time-period of a cardiac cycle. We report the minimum and maximum TAWSS and OSI, respectively, within the stenotic region of each of the three primary coronary arteries.

Note that 0D low-fidelity models cannot capture spatially-resolved flow-fields, and therefore can only provide coarse approximations of TAWSS and OSI based on analytic solutions. Therefore, we used the mean flow at each outlet as the low-fidelity predictor ($\Qlf$) for both quantities.
\begin{figure}
\centering
\includegraphics[scale=1.0]{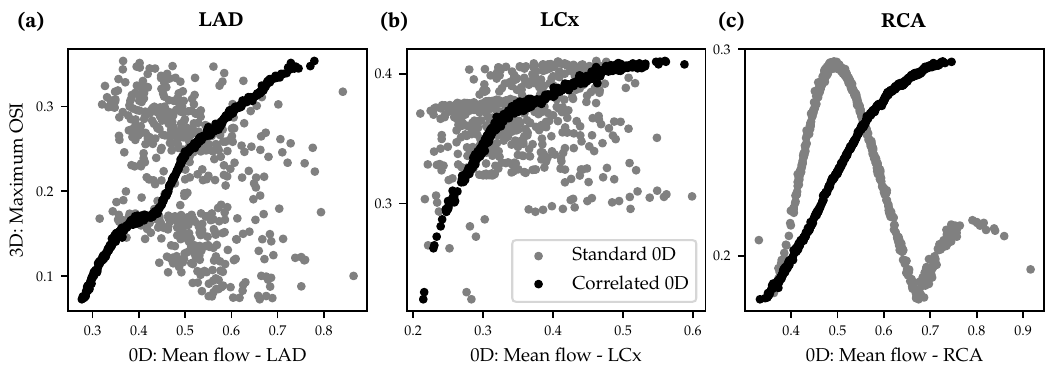}
\caption{Correlation between maximum OSI from 3D simulations and mean outlet flow from 0D simulations. Data is shown for the standard and MFMC-AE-reparameterized 0D models. The three panels show the correlations for the LAD, LCx and RCA branches.}
\label{fig:max_osi_scatter}
\end{figure}
Figure~\ref{fig:max_osi_scatter} shows the correlation between the maximum 3D model OSI for the three primary coronary arteries versus the corresponding mean flow computed by the 0D model, where MFMC-AE produces remarkably higher correlations.
\begin{figure}[h]
\centering
\includegraphics[scale=1.0]{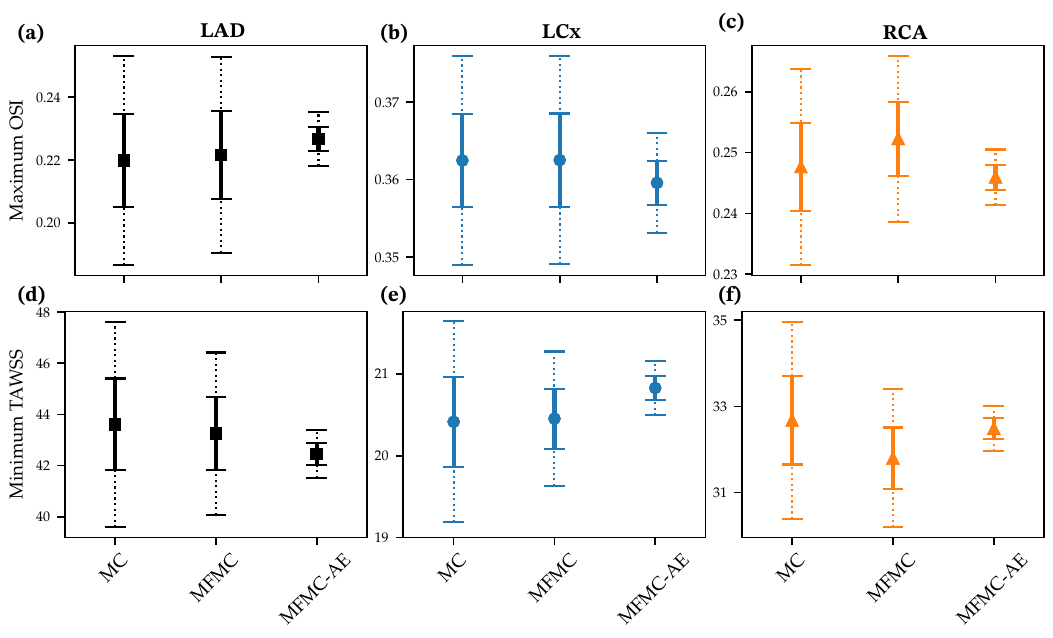}
\caption{Maximum OSI in (a)-(c) and minimum TAWSS in (d)-(f) for LAD, LCx and RCA branches estimated using one realization of MC, MFMC and MFMC-AE estimators. Solid and dashed lines show 95\% and 99\% confidence intervals, respectively.}
\label{fig:qoi_wss}
\end{figure}

The Pearson correlation coefficient between the maximum OSI and the mean flow was -0.3635 for the LAD, 0.0998 for the LCx, and -0.5493 for the RCA. The current framework improved these correlations to 0.9906 for the LAD, 0.9018 for the LCx, and 0.9840 for the RCA. Error-bars with 95\% and 99\% confidence intervals for the maximum OSI in each of the three branches are shown in Figures~\ref{fig:qoi_wss}(a)-(c) for the MC, MFMC and MFMC-AE estimators, respectively. The proposed MFMC-AE estimator achieves a roughly three-fold reduction in the variance.

For the case of minimum TAWSS, the Pearson correlation coefficient between the minimum TAWSS from 3D simulations ($\Qhf$) and the 0D-estimated mean flow ($\Qlf$) was 0.6294 for the LAD, 0.7622 for the LCx, and 0.7349 for the RCA. The current framework improved this to 0.9977 for the LAD, 0.9873 for the LCx, and 0.9979 for the RCA. Figures~\ref{fig:qoi_wss}(d)-(f) show estimates of minimum TAWSS from the MC, MFMC, and MFMC-AE estimators for the three coronary branches. In this case too, we observed a three-fold improvement in the variance of the estimator.

\subsubsection{Fractional flow reserve}
\label{sec:ffr_imr}

\begin{figure}
\centering
\includegraphics[scale=1.0]{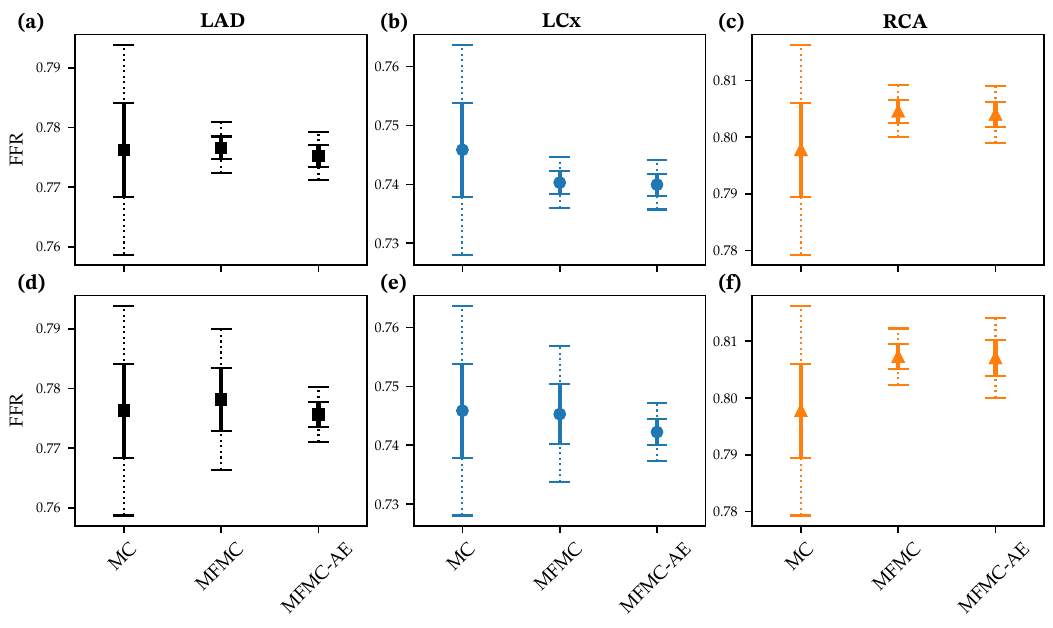}[h]
\caption{FFR estimated with 0D FFR in (a)-(c), and 0D mean flow in (d)-(f), using one realization of MC, MFMC and MFMC-AE. The three panels show data for the LAD, LCx and RCA branches. Solid and dashed lines show 95\% and 99\% confidence intervals.}
\label{fig:qoi_ffr}
\end{figure}

Catheter-based measurement of fractional flow reserve (FFR) is the current clinical gold-standard for the assessment of functional CAD severity and decision-making regarding coronary interventions~\citep{Pijls1996}. FFR is measured as the ratio of the pressure immediately distal to a lesion divided by the aortic pressure under hyperemia. Consistent with this definition, we computed FFR from 3D simulations as the ratio of the cross-section-average pressure downstream of the stenotic region for each of the three primary vessels, divided by the aortic pressure. 

The computation of FFR from 0D simulations is generally not possible in an analogous manner because these models do not include a spatial description of the flow and anatomy. Therefore, there is generally no output that is equivalent to the pressure at an anatomical location immediately distal to a stenosis. Moreover, these models usually assume that pressure loss along vessels is linear, inspired by Poiseuille flow. We included two specific features to address this. Each length of artery, separated by bifurcations, was modeled as a distinct vessel. This allowed us to isolate the location of each stenosis more precisely. We also modeled the non-linear pressure loss due to vascular stenoses as $\Delta P = S|Q|Q$, where $\Delta P$ is the pressure loss, $Q$ is the flow rate, and $S$ is a coefficient determined primarily by the degree of stenosis \citep{Pfaller2022AutomatedFlow}. These feature allowed us to compute FFR from the 0D simulations in a similar manner to its computation from the 3D simulations. However, to demonstrate the efficacy of the current uncertainty quantification technique when using more standard (and widely used) 0D models that do not include these specialized features, we also report results where the low-fidelity estimator is simply the mean flow in each branch.

The Pearson correlation coefficient between FFR computed from the 3D and 0D simulations was 0.9951 for the LAD, 0.9954 for the LCx, and 0.9940 for the RCA. This high correlation is the result of the 0D model features described above. The current MFMC-AE framework was able to maintain similar correlations, specifically 0.9983 for the LAD, 0.9967 for the LCx, and 0.9868 for the RCA. Figures~\ref{fig:qoi_ffr}(a)-(c) show estimated FFR with 95\% and 99\% confidence intervals using MC, MFMC, and MFMC-AE estimators. We observed significant improvement in the confidence intervals with both MFMC and MFMC-AE, compared to MC.

When using the 0D mean flow as as low-fidelity predictor for FFR, the Pearson correlation between the 3D and 0D outputs dropped to -0.7612 for the LAD, -0.7820 for the LCx, and -0.9884 for the RCA. As expected, the mean flow was a less adequate low-fidelity surrogate for FFR, especially for the LAD and LCx. In this case, MFMC-AE was able to restore correlations to 0.9901 for the LAD, 0.9853 for the LCx, and 0.9495 for the RCA. The improved estimate for FFR in the LAD and LCx is evident in the confidence intervals shown in Figures~\ref{fig:qoi_ffr}(d)-(f).

A summary of all the computed quantities, the correlation between low and high-fidelity models, and the standard deviation of the three estimators is provided in Table~\ref{table:UQ}. 
\begin{table}
\centering
\begin{tabular}{ |c|c|c|c|c|c|c|c| } 
\hline
\multicolumn{2}{|c|}{\textbf{Quantity of interest}} & \multirow{2}{4em}{\textbf{Branch}} & \multicolumn{2}{|c|}{\textbf{Correlation}} & \multicolumn{3}{|c|}{\textbf{Standard deviation}} \\ \cline{1-2} \cline{4-8}
\textbf{3D} & \textbf{0D} & & \textbf{Original} & \textbf{Modified} & \textbf{MC} & \textbf{MFMC} & \textbf{MFMC-AE} \\
\hline
\hline
\multirow{3}{4em}{Max. OSI} & \multirow{3}{4em}{Mean flow} & LAD & -0.3635 & 0.9906 & 0.0033 & 0.0031 & 0.0009 \\ 
& & LCx & 0.0998 & 0.9018 & 0.0013 & 0.0013 & 0.0006 \\ 
& & RCA & -0.5493 & 0.9840 & 0.0016 & 0.0014 & 0.0005 \\ 
\hline
\hline
\multirow{3}{4em}{Min. TAWSS} & \multirow{3}{4em}{Mean flow} & LAD & 0.6294 & 0.9977 & 0.4017 & 0.3173 & 0.0936 \\ 
& & LCx & 0.7622 & 0.9873 & 0.1228 & 0.0822 & 0.0334 \\ 
& & RCA & 0.7349 & 0.9979 & 0.2289 & 0.1597 & 0.0532 \\ 
\hline
\hline
\multirow{3}{4em}{FFR} & \multirow{3}{4em}{FFR} & LAD & 0.9951 & 0.9983 & 0.0018 & 0.0004 & 0.0004 \\ 
& & LCx & 0.9954 & 0.9967 & 0.0018 & 0.0004 & 0.0004 \\ 
& & RCA & 0.9940 & 0.9868 & 0.0018 & 0.0005 & 0.0005 \\ 
\hline
\hline
\multirow{3}{4em}{FFR} & \multirow{3}{4em}{Mean flow} & LAD & -0.7612 & 0.9901 & 0.0018 & 0.0012 & 0.0005 \\ 
& & LCx & -0.7820 & 0.9853 & 0.0018 & 0.0012 & 0.0005 \\ 
& & RCA & -0.9884 & 0.9495 & 0.0018 & 0.0005 & 0.0007 \\ 
\hline
\end{tabular}
\caption{Summary of the computed quantities of interest.}
\label{table:UQ}
\end{table}

\subsection{Computational budget}
\label{sec:budget}

\begin{figure}[h]
\centering
\includegraphics[scale=1.0]{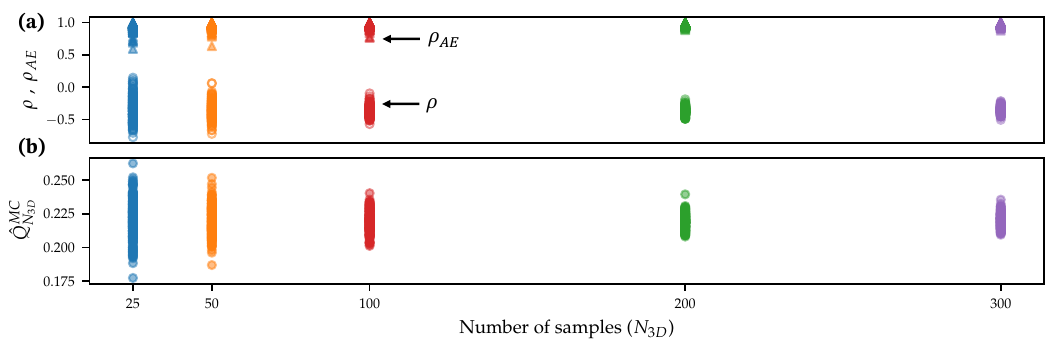}
\caption{Convergence of MFMC-AE with respect to $\Nhf$ using maximum OSI as the QoI. Top panel shows original and modified correlations between 3D and 0D outputs. Bottom panel shows the Monte Carlo mean. Each marker is an independent trial.}
\label{fig:convergence}
\end{figure}
We now discuss the computational cost of the proposed framework, using the maximum OSI as an example QoI. We discuss the convergence of the framework with respect to the pilot sample $\Nhf$, as well as the cost of estimating QoIs using the current MFMC-AE approach compared with standard MC and MFMC techniques. 

We begin with a discussion of the number of 3D samples required for convergence. As noted earlier, we used $\Nhf = 500$ high-fidelity simulations in this work. To assess convergence for lower values of $\Nhf$, we sub-sampled from the original $\Nhf = 500$ pilot sample for the inputs and performed 200 independent trials of the MFMC-AE framework for each $\Nhf$. Figure~\ref{fig:convergence} shows the convergence of the framework for increasing $\Nhf$ in terms of the original correlation ($\rho$) between 3D and 0D outputs, the modified correlation ($\rho_{AE}$), and the Monte Carlo estimate of the mean ($\Qmc_\Nhf$). Remarkably, the framework is able to increase the correlation between 0D and 3D evaluations even for as low as $\Nhf = 25$. However, the repeatability between independent trials significantly improves for increasing $\Nhf$ and converges for approximately $\Nhf=100$. 

We now analyze the computational budget for the estimation of QoIs using the standard MC and MFMC techniques, compared to the current MFMC-AE framework. 
Note that the computational cost is reported in terms of equivalent HF simulations by dividing the total cost by the cost for a single 3D simulation (see Section~\ref{sec:modeling}). 

\begin{figure}
\centering
\includegraphics[scale=1.0]{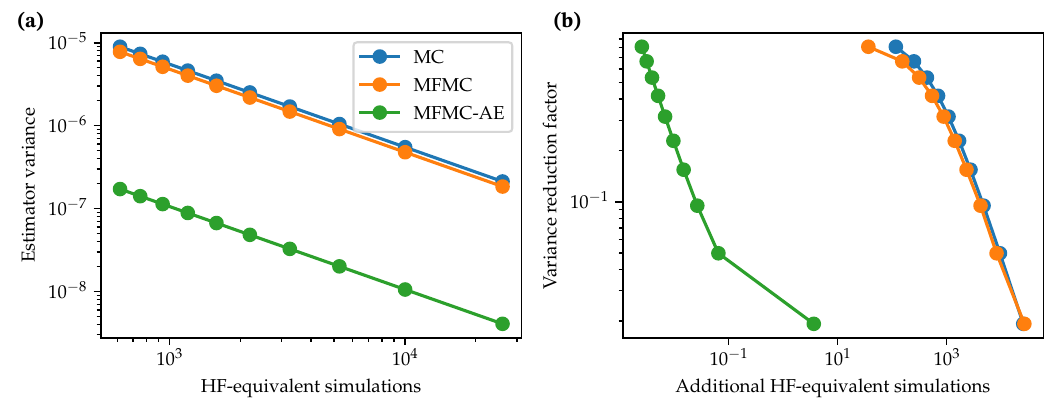}[h]
\caption{Computational budget analysis using maximum OSI as an example QoI. (a) Optimal variance reduction versus computational cost. (b) Additional computational budget required to reduce current (non-optimal) estimator variance.}
\label{fig:cost}
\end{figure}

Figure~\ref{fig:cost}(a) compares the optimal/minimum variance of the MC, MFMC and MFMC-AE estimators as a function of the computational budget. For a given computational budget, the optimal variance was obtained following the work of Peherstorfer \textit{et al.} \citep{Peherstorfer2016OptimalEstimation}. 
The optimal number of high-fidelity and low-fidelity simulations for a given budget, $\mathcal{B}$ is,
\begin{align}
    \Nhf = \frac{\mathcal{B}}{1+w\gamma}  \quad ; \quad \Nlf = \gamma \Nhf \quad ; \quad \gamma = \sqrt{\frac{\rho^2}{w(1-\rho^2)}},
    \label{eq:optimal_allocation}
\end{align}
where $w \ll 1$ represents the ratio of low-fidelity to high-fidelity model computational cost. The optimal variance for the MFMC and MFMC-AE estimators is given by using Equation~\eqref{eq:optimal_allocation} in Equation~\eqref{eq:mfmc_var}. Figure~\ref{fig:cost}(a) shows a reduction in estimator variance of roughly $\mathcal{O}(10^{-2})$ resulting from the MFMC-AE estimator. 

We note that in this work, $\Nlf=10,000$. 
This is approximately 50 times fewer low-fidelity evaluations than the optimal number suggested by Equation~\ref{eq:optimal_allocation}. We therefore also report the incremental value of increasing the current computational budget to reduce the variance of the current estimate of the maximum OSI (Figure~\ref{fig:cost}(b)). We obtain an order of magnitude reduction in the estimator variance for $\sim$10\% of the cost of a single three-dimensional simulation. 
On the other hand, the MC and MFMC methods would require over $10^3$ additional 3D simulations.

\section{Discussion}

This paper provides a proof-of-concept demonstration of a novel uncertainty-aware framework for simulating patient-specific and vessel-specific coronary artery flows, informed by non-invasive \mpi and routine measurements of cardiac function. The main outcomes of this work were two-fold. First, we developed an end-to-end pipeline to estimate model parameters with the ability to match measured cardiac function and coronary flows, and to predict clinical and biomechanical quantities under this data-informed uncertainty. Second, we demonstrated significantly reduced variance and computational cost in the prediction of QoIs using a novel strategy for multi-fidelity uncertainty propagation relying on non-linear dimensionality reduction.

This study is motivated by prior work by Menon \textit{et al.}~\cite{Menon2024} which highlighted the need for personalized vessel-specific flows in modeling coronary artery hemodynamics. Their work showed significant differences in coronary flows as well as FFR predicted by models that distribute the flow amongst coronary arteries based on patient-specific myocardial blood flow distributions from \mpi versus empirical diameter-based rules. The incorporation of \mpi into simulations of coronary flow is especially promising for clinical translation because current image-based coronary flow models are also based on CT imaging. Therefore, the high-resolution quantification of myocardial blood flow and full coverage of the myocardium from \mpi can potentially be incorporated into existing CT-based imaging protocols -- streamlining the combination of these imaging techniques to develop more personalized hemodynamics simulations. In fact, the quantification of patient-specific hemodynamics through either simulations or \mpi have already been shown to provide improved clinical value in the diagnosis and treatment of CAD compared to traditional anatomical imaging via CT or invasive angiography alone \citep{Pijls2010,Nous2022DynamicDisease,TAYLOR2023116414}. Therefore, we expect increased personalization via the combination of novel imaging and modeling techniques to improve the accuracy and clinical value of these tools.

The inclusion of hemodynamic measurements in clinical CAD risk stratification is especially valuable in lesions that have borderline severity \citep{Tonino2009,Nakazato2013NoninvasiveStudy}. Therefore, for the robust clinical translation of these simulation-based techniques, it is imperative to assess confidence in the predicted clinical risk. To that end, we extended our previous work on deterministic coronary flow personalization~\citep{Menon2024} to account for uncertainty in the clinical data.

While previous work involving uncertainty quantification for cardiovascular hemodynamics has either focused on the estimation of uncertain model parameters or the prediction of uncertain quantities of interest, here we demonstrated an end-to-end pipeline that goes from the clinical imaging to predicted clinical and biomechanical QoIs. For the estimation of personalized model parameters, we combined deterministic optimization for the parameters governing cardiac function with Bayesian estimation for the parameters that dictated vessel-specific coronary flows. Using a DREAM MCMC sampler~\citep{Vrugt2009AcceleratingSampling}, combining parallel Markov chains with differential evolution and adaptive proposal distributions, we demonstrated the ability to sample from a relatively high-dimensional posterior distribution with 14 input parameters. This method is also known to show improvement over other MCMC techniques when sampling from heavy-tailed or multi-modal posterior distributions. 

We also demonstrated a novel framework to significantly improve the confidence in predictions from simulations that incorporate noisy clinical data. While multi-fidelity Monte Carlo estimators are a mainstay for applications which can leverage computationally inexpensive low-fidelity surrogate models to reduce the variance of quantities estimated by expensive high-fidelity simulations \citep{Peherstorfer2016OptimalEstimation}, they rely on the availability of correlated low-fidelity surrogates. However, this is not always the case -- especially for applications that aim to capture non-linear physics that is difficult to represent in low-fidelity models. We highlighted this deficiency in the estimation of TAWSS and OSI, which are governed by non-linear flow physics in the vicinity of the stenosis where their effect is most useful to estimate. 

Using non-linear dimensionality reduction, we modified the sampling for existing low-fidelity 0D models to significantly improve their correlation with respect to high-fidelity 3D models. This was achieved by creating a \textit{shared space} between the models -- a one-dimensional non-linear manifold in the input space of each model that could capture most of the variability in the output QoI, and where the one-dimensional representation of the inputs for each model had the same distribution (the standard Gaussian). We then transformed each parameter vector for the 3D model to the shared space, and reconstructed the corresponding 0D input vector for each 0D evaluation. Intuitively, we ensured that the low-fidelity model was evaluated at the same locations as the high-fidelity model along the axes which are most influential for each of them, thus increasing their correlation.


Through the novel combination of multi-fidelity Monte Carlo with non-linear dimensionality reduction demonstrated here, we showed significant improvements in the confidence intervals for our predictions for both biomechanical as well as clinical quantities of interest. We also showed that for cases when we already have correlated low-fidelity surrogate models, such as for the FFR computation which included a model for the non-linear pressure drop within stenoses, our method preserves the effectiveness of the standard multi-fidelity Monte Carlo technique. Finally, we achieved significant variance reduction in the multi-fidelity Monte Carlo estimator with orders of magnitude reduction in the computational cost. This was demonstrated for the optimal number of simulations as well as with a more practical smaller computational budget. 

While this study demonstrates a promising direction in uncertainty-aware modeling of personalized coronary artery flows, there are several limitations and future directions that deserve attention. First, while the current work represents a proof-of-concept, the methods presented here should be evaluated in large patient populations to demonstrate their clinical value. Moreover, we only considered the effect of uncertainty in \mpi. While the main focus of this work was on vessel-specific coronary flows, the personalization of the model will be influenced by uncertainty in all the other clinical measurements utilized, including anatomical imaging, echocardiography, etc. Future work should evaluate the sensitivity of the computational modeling to these myriad sources of clinical uncertainty to find the most influential effects. Moreover, due to the absence of repeated clinical measurements, this work used Gaussian noise to simulate clinical uncertainty in \mpi. However, robust clinical applications would benefit from clinically evaluated variability (such as inter- and intra-operator variability) in the measurements.

The inclusion of additional sources of uncertainty would likely lead to very high-dimensional stochastic inputs. Extensions of this work would therefore benefit from recent developments in the area of Bayesian estimation, such as data-driven simulation-based inference techniques which combine variational inference with identifiability and sensitivity analysis for high-dimensional problems \cite{Cranmer2020TheInference,TONG2024116846}. In addition, while we used the zero-dimensional model as a surrogate for the parameter estimation, as has been done in previous work \citep{Tran2017,Menon2023,Menon2024}, multi-fidelity approaches would improve the accuracy of the parameter estimation \citep{richter2024bayesianwindkesselcalibrationusing}. Although not demonstrated here, our novel multi-fidelity Monte Carlo technique allows the use of high-fidelity and low-fidelity models which have dissimilar input parameters \citep{ZANONI2024117119}. This is enabled by the construction of the shared reduced-dimensional space between the models. Therefore, future work should explore promising avenues to further reduce computational cost by utilizing pre-computed simulation libraries with dissimilar inputs as surrogate models for relevant quantities of interest. 

\section{Conclusions}

We demonstrated an end-to-end pipeline for personalized prediction of coronary hemodynamics under clinical data uncertainty. We utilized novel and routine clinical measurements -- of myocardial blood flow from CT myocardial perfusion imaging and cardiac function from echocardiography -- to personalize computational models of coronary hemodynamics in terms of gross hemodynamics as well as vessel-specific coronary flows. To account for uncertainty in the clinical data, Bayesian estimation was performed on personalized model parameters using adaptive Markov chain Monte Carlo. By combining multi-fidelity Monte Carlo uncertainty propagation with data-driven non-linear dimensionality reduction, we determined clinical and biomechanical posterior predictive QoIs, under uncertainty stemming from the clinical measurements. We demonstrated significantly improved correlations with respect to reference multi-fidelity estimators. For relevant quantities of interest, we showed that improved correlations greatly reduces estimator variance compared to standard multi-fidelity estimators, offering orders-of-magnitude computational cost savings.

\section{Data availability}

All computational models built for this study have been anonymized and made publicly available through the Vascular Model Repository \textit{(https://vascularmodel.com)}.

\section{Code availability}

The patient-specific modeling pipeline and computational fluid dynamics solvers are part of the SimVascular open-source project \textit{(https://simvascular.github.io/)}. Image analysis and postprocessing was performed using the open-source tools specified in section \ref{sec:clinical_data}. The software to perform the non-linear dimensionality reduction and uncertainty quantification is available on Github at \textit{https://github.com/StanfordCBCL/NeurAM}.

\section{Acknowledgments}

This work was supported by NIH grant R01HL141712, NSF CAREER award 1942662, and NSF CDS\&E awards 2104831 and 2105345. High performance computing resources were provided by the Stanford Research Computing Center. Sandia National Laboratories is a multi-mission laboratory managed and operated by National Technology \& Engineering Solutions of Sandia, LLC (NTESS), a wholly owned subsidiary of Honeywell International Inc., for the U.S. Department of Energy’s National Nuclear Security Administration (DOE/NNSA) under contract DE-NA0003525. This written work is authored by an employee of NTESS. The employee, not NTESS, owns the right, title and interest in and to the written work and is responsible for its contents. Any subjective views or opinions that might be expressed in the written work do not necessarily represent the views of the U.S. Government. The publisher acknowledges that the U.S. Government retains a non-exclusive, paid-up, irrevocable, world-wide license to publish or reproduce the published form of this written work or allow others to do so, for U.S. Government purposes. The DOE will provide public access to results of federally sponsored research in accordance with the DOE Public Access Plan. 

\section*{Appendix}
\setcounter{figure}{0}
\renewcommand\thefigure{\thesection.\arabic{figure}}
\appendix

\section{Neural network architecture and hyperparameters}
\label{app:hyperparameters}

The neural networks for both the autoencoders ($\encoderhf$, $\decoderhf$, $\encoderlf$, $\decoderlf$) were parameterized as fully-connected neural networks. The encoders and decoders both had 5 layers with 13 neurons each. The first and fifth layers were linear layers, and the other layers had hyperbolic tangent activation functions. The decoder had an additional sigmoidal layer as its last layer to ensure that the reconstructed inputs lay within the input parameter space. 

The neural network surrogates for both the high- and low-fidelity quantities of interest, $\Qhf_{NN}$ and $\Qlf_{NN}$, were fully connected networks with 3 layers and 10 neurons each. The first and last layer were linear, while the middle layer used a hyperbolic tangent activation.

All the neural networks were trained using the Adam optimizer \citep{kingma2017} and an exponential scheduler for the learning rate. Hyper-parameters were tuned using the \textit{Optuna} \citep{optuna_2019}.

\section{Details on parameter estimation}
\label{app:parameter_estimation}
\begin{figure}
\centering
\includegraphics[scale=1.0]{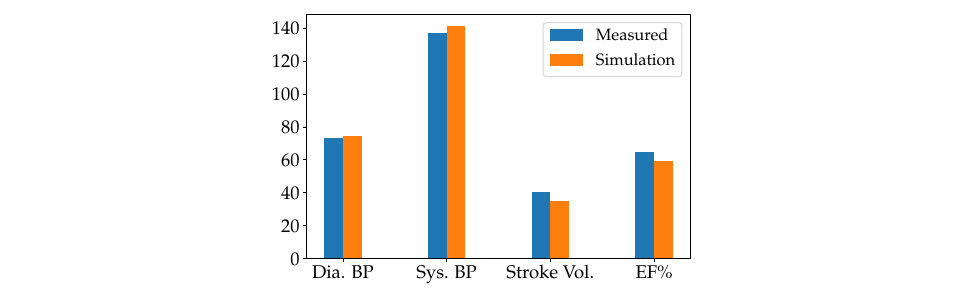}
\caption{Comparison of measured versus simulated metrics of cardiac function after model personalization. Dia. BP: Diastolic blood pressure. Sys. BP: Systolic blood pressure. Stroke Vol.: Stroke volume. EF: Ejection fraction.}
\label{fig:cardiac_function}
\end{figure}
As a first step of the parameter estimation process we used deterministic optimization to tune the parameters of the closed-loop lumped-parameter network. We used gradient-free Nelder-Mead optimization, with the primary targets being measurements of cardiac function from echocardiography and blood pressure cuff measurements. Further details on this parameter tuning are available in Menon \textit{et al.} \citep{Menon2024}. Figure \ref{fig:cardiac_function} shows the results of this tuning for the patient-specific data used in this study. We see that we successfully recapitulated the measurements of blood pressure as well as stroke volume and ejection fraction.

\begin{figure}
\centering
\includegraphics[scale=0.3]{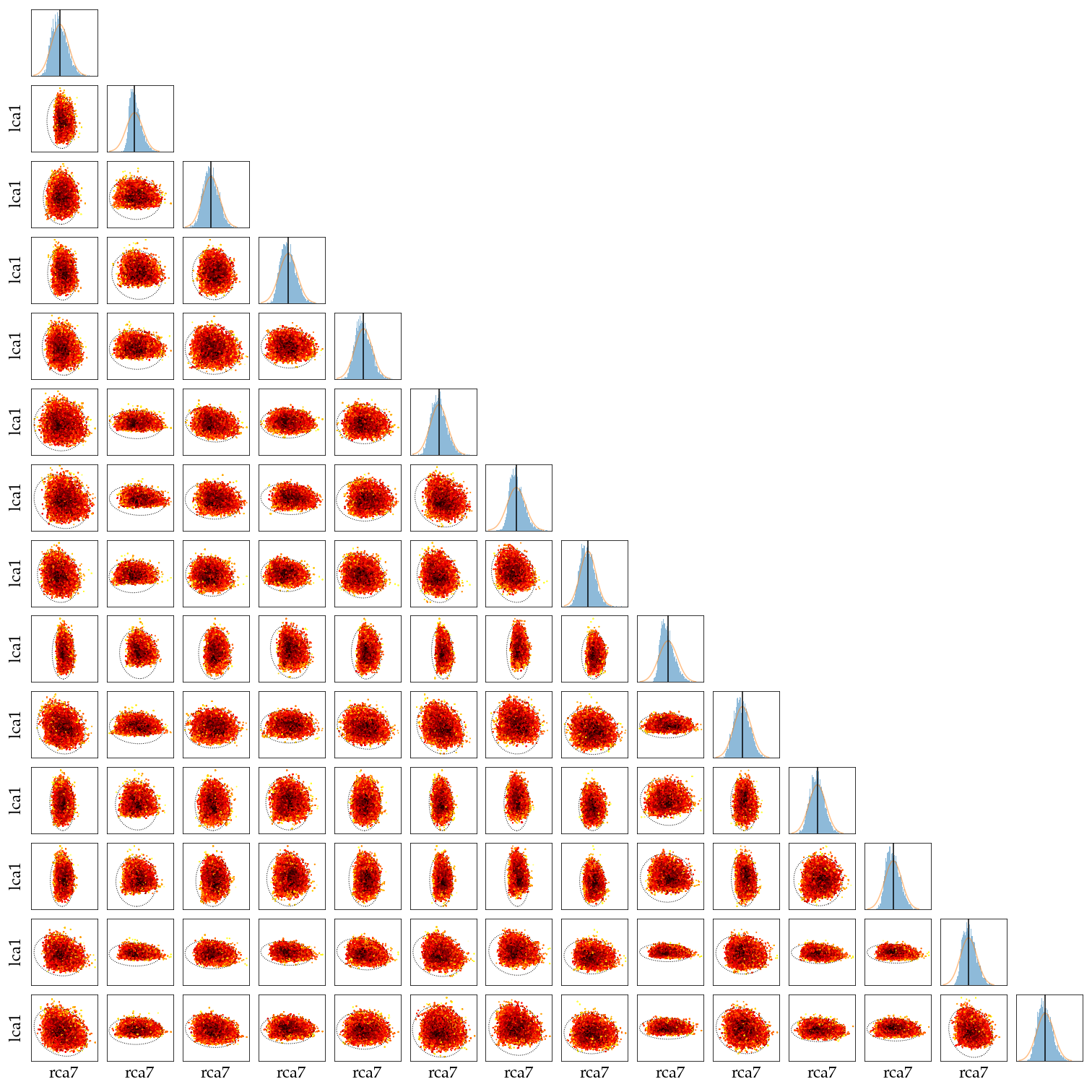}
\caption{Distributions of simulated (scatter) versus measured (dashed ellipse) covariance between branch-specific coronary flows resulting from Bayesian parameters estimation.}
\label{fig:results_scatter}
\end{figure}
When extracting the clinical targets for the second stage of the parameter estimation, i.e. the branch-specific coronary flows with simulated Gaussian noise, we also obtained the covariance of the targets ($\boldsymbol{\Sigma}$). This covariance was a result of the interdependence amongst nearby branches when partitioning of the LV myocardium into non-overlapping perfusion territories corresponding to each coronary artery branch. The covariance was included in the Bayesian parameter estimation, as shown in the likelihood function in Equation~\eqref{eq:likelihood}. For each pair of coronary branches, Figure \ref{fig:results_scatter} compares the predicted covariance from the parameter estimation, shown as scatter plots, with the target covariance, shown as ellipses. We see that the parameter estimation was able to capture the target covariance reasonably well.


\bibliographystyle{unsrt}

\end{document}